\documentclass[10pt,conference]{IEEEtran}
\IEEEoverridecommandlockouts

\usepackage{cite}
\usepackage{amsmath,amssymb,amsfonts}
\usepackage{algorithmic}
\usepackage{graphicx}
\usepackage{textcomp}
\usepackage{xcolor}
\usepackage{subcaption}
\usepackage{braket}
\usepackage{tikz}
\usepackage{qcircuit}
\usepackage{booktabs} 
\usepackage{tabularx}
\usepackage{caption}
\usepackage{atbegshi}
\usepackage{eso-pic}
\AtBeginShipout{%
  \ifnum\value{page}=4
\global\setbox\AtBeginShipoutBox=\vbox{\vspace{2pt}\unvbox\AtBeginShipoutBox}%
  \fi
}
\def\BibTeX{{\rm B\kern-.05em{\sc i\kern-.025em b}\kern-.08em
T\kern-.1667em\lower.7ex\hbox{E}\kern-.125emX}}

\begin{document}

\title{Benchmarking Quantum and Classical Sequential Models for Urban Telecommunication Forecasting \thanks{The views expressed in this article are those of the authors and do not represent the views of Wells Fargo. This article is for informational purposes only. Nothing contained in this article should be construed as investment advice. Wells Fargo makes no express or implied warranties and expressly disclaims all legal, tax, and accounting implications related to this article. }}

\author{
\IEEEauthorblockN{Chi-Sheng Chen}
\IEEEauthorblockA{
\textit{Neuro Industry, Inc.}\\
Cambridge, Massachusetts, USA \\
m50816m50816@gmail.com}
\and
\IEEEauthorblockN{Samuel Yen-Chi Chen}
\IEEEauthorblockA{\textit{Wells Fargo} \\
New York, USA \\
yen-chi.chen@wellsfargo.com}
\and
\IEEEauthorblockN{Yun-Cheng Tsai}
\IEEEauthorblockA{\textit{National Taiwan Normal University}\\
Taipei, Taiwan \\
pecu@ntnu.edu.tw}
}

\maketitle

\begin{abstract}
In this study, we evaluate the performance of classical and quantum-inspired sequential models in forecasting univariate time series of incoming SMS activity (SMS-in) using the Milan Telecommunication Activity Dataset. Due to data completeness limitations, we focus exclusively on the SMS-in signal for each spatial grid cell. We compare five models—LSTM (baseline), Quantum LSTM (QLSTM), Quantum Adaptive Self-Attention (QASA), Quantum Receptance Weighted Key-Value (QRWKV), and Quantum Fast Weight Programmers (QFWP), under varying input sequence lengths (4, 8, 12, 16, 32 and 64). All models are trained to predict the next 10-minute SMS-in value based solely on historical values within a given sequence window. Our findings indicate that different models exhibit varying sensitivities to sequence length, suggesting that quantum enhancements are not universally advantageous. Rather, the effectiveness of quantum modules is highly dependent on the specific task and architectural design, reflecting inherent trade-offs among model size, parameterization strategies, and temporal modeling capabilities.

\end{abstract}


\section{Introduction}
Time series forecasting is a critical component in urban informatics, enabling efficient resource allocation, network optimization, and proactive management in telecommunication systems. Urban telecommunication data, such as call volumes, SMS activity, and internet usage, exhibit complex temporal patterns influenced by human mobility, events, and daily routines. Accurate prediction of these signals can support applications in smart city planning, traffic management, and emergency response \cite{barlacchi2015multi}. The Milan Telecommunication Activity Dataset, which captures multi-source urban life indicators including telecommunication activity across a spatial grid in Milan, Italy, serves as a valuable benchmark for such forecasting tasks \cite{barlacchi2015multi}.

Classical sequential models, particularly recurrent neural networks (RNNs) like Long Short-Term Memory (LSTM), have been the cornerstone for time series prediction due to their ability to capture long-range dependencies in sequential data \cite{hochreiter1997long}. However, these models often struggle with vanishing gradients and computational inefficiency in very long sequences. Recent advancements in quantum machine learning (QML) promise to address these limitations by leveraging quantum superposition and entanglement for enhanced representational power and potentially faster processing \cite{biamonte2017quantum}.

In this context, quantum-inspired architectures have emerged as hybrid solutions that integrate quantum principles into classical frameworks. For instance, Quantum Long Short-Term Memory (QLSTM) extends the classical LSTM by incorporating quantum gates for improved memory mechanisms \cite{chen2020quantum}. Similarly, Quantum Adaptive Self-Attention (QASA) introduces a quantum-enhanced attention module to classical Transformers, enabling adaptive focus on relevant sequence elements \cite{chen2025qasa}. Quantum Receptance Weighted Key-Value (QRWKV) builds on the Receptance Weighted Key-Value (RWKV) model, infusing quantum operations to optimize token interactions in linear time \cite{peng2023rwkv, chen2025qrwkv}. Finally, Quantum Fast Weight Programmers (QFWP) utilize quantum fast weights for the dynamic programming of variational quantum circuits, enabling rapid adaptation to sequential patterns \cite{chen2024qfwp}.

However, several challenges arise when applying sequential models to such data. First, the dataset suffers from incomplete or sparse entries across many modalities, limiting the feasibility of multivariate modeling. Second, the temporal dynamics of urban telecommunication signals can be highly non-stationary and vary across spatial regions. Third, determining the optimal input sequence length for forecasting remains a nontrivial task, as different models may exhibit varying sensitivities to short- or long-term dependencies.

To address these challenges, this paper focuses on a univariate forecasting task: predicting the next SMS-in value for each grid cell based on a fixed-length window of past SMS-in values. We compare the performance of five models: a classical Long Short-Term Memory (LSTM) network as the baseline, and four quantum-inspired models: Quantum Adaptive Self-Attention (QASA), Quantum LSTM (QLSTM), Quantum Receptance Weighted Key Value (QRWKV), and Quantum Fast Weight Programmers (QFWP). We systematically evaluate each model across different input sequence lengths (4, 8, 12, 16, 32 and 64), revealing how model performance varies with temporal context.

Our contributions are threefold:
\begin{enumerate}
    \item We construct a univariate forecasting benchmark using the Milan telecommunication dataset, focusing on SMS-in signals while considering practical constraints from real-world data sparsity.
    \item We provide a comprehensive comparison of classical and quantum sequential models in a real-world urban telecommunication scenario, including insights into model robustness and sequence-length sensitivity under different sequence windows.
    \item We demonstrate that specific quantum models (e.g., QASA, QRWKV, QFWP and QLSTM) exhibit distinct advantages in capturing short- or long-range dependencies, offering empirical evidence on the practical benefits of quantum enhancements for time series tasks under limited data settings.
\end{enumerate}

\section{Related Work}
\subsection{Time Series Forecasting Models}
Time series forecasting is a foundational task across various domains, including finance, energy, and urban infrastructure. Traditional statistical models, including the Autoregressive Integrated Moving Average (ARIMA)~\cite{box1970time}, Holt-Winters exponential smoothing~\cite{winters1960forecasting}, and other exponential smoothing techniques~\cite{holt1957forecasting}, have been extensively employed due to their simplicity and interpretability. However, these methods often fall short in modeling nonlinear temporal dependencies and handling non-stationary data effectively~\cite{hyndman2008forecasting}. The rise of deep learning has shifted the paradigm toward models that can capture intricate sequence patterns. Recurrent Neural Networks (RNNs), especially Long Short-Term Memory (LSTM) networks~\cite{hochreiter1997long}, have become benchmarks due to their gated mechanisms, which mitigate vanishing gradients and enable the modeling of long-range dependencies. More recently, attention-based architectures, such as the Transformer~\cite{vaswani2017attention} and its specialized variants for time series, such as Informer~\cite{zhou2021informer} and Autoformer~\cite{wu2021autoformer}, have demonstrated superior performance in long-horizon forecasting by leveraging self-attention to capture global interactions without requiring sequential processing.

\subsection{Quantum-Inspired Sequence Models}
Quantum machine learning introduces innovative paradigms that emulate quantum phenomena, such as superposition and entanglement, in classical systems, offering novel inductive biases for sequence modeling. These approaches are particularly promising for tasks that require the efficient capture of complex dependencies and adaptation in data-scarce environments. Quantum Long Short-Term Memory (QLSTM) enhances the classical LSTM by integrating trainable quantum circuits to improve memory retention and dynamics~\cite{chen2020quantum}. Quantum Receptance Weighted Key-Value (QRWKV) extends the RWKV framework~\cite{peng2023rwkv} with quantum operations, enabling linear-time token interactions through recurrently updated representations that simulate quantum memory decay~\cite{chen2025qrwkv}. Quantum Adaptive Self-Attention (QASA) incorporates unitary-inspired computations into attention mechanisms, facilitating adaptive context encoding with enhanced parameter efficiency and gradient propagation~\cite{chen2025qasa}. Quantum Fast Weight Programmers (QFWP) harness the synergy between associative memory and variational quantum circuits. By computing fast weights as outer products of deep neural network outputs, QFWP dynamically generate context-sensitive quantum neural network parameters, facilitating rapid adaptation to evolving sequences~\cite{chen2024qfwp}. These models not only address computational bottlenecks in classical architectures but also show potential for superior generalization in low-resource settings, as evidenced by recent applications in time series forecasting tasks~\cite{chen2020quantum,chen2024qfwp}.

\subsection{QML Cross-Sector Application }
QML has also been explored in \emph{finance}, \emph{energy}, and \emph{transportation}, showing benefits in analytics, and optimization enhancement. These examples highlight the broader impact of QML and motivate our focus on hybrid quantum–classical \emph{sequence models} tailored to telecom time-series forecasting.

\subsection{Telecommunication Activity Prediction}
Telecommunication activity prediction is vital for urban planning, mobility analysis, and infrastructure optimization. The Milan Telecommunication Activity Dataset~\cite{barlacchi2015multi} offers a comprehensive resource for such studies, encompassing spatiotemporal records of SMS, calls, and internet traffic in Milan, Italy. Prior research has utilized this dataset for tasks like population density estimation~\cite{deville2014dynamic}, spatial clustering and hotspot detection~\cite{geertman2021urban}, and multivariate traffic forecasting~\cite{barlacchi2015multi}. For instance, studies have explored dynamic population mapping using call detail records~\cite{deville2014dynamic} and anomaly detection in urban activity patterns~\cite{geertman2021urban}. However, challenges such as data sparsity and incompleteness across modalities have limited the effectiveness of multivariate approaches, often leading to a focus on aggregated or specific signals.

Despite these advancements, a gap remains in systematically benchmarking quantum-inspired sequential models, such as QLSTM, QASA, QRWKV, and QFWP, against classical baselines on real-world urban datasets, particularly when varying input sequence lengths are considered. While recent works have applied quantum machine learning to related domains, such as metropolitan traffic time series forecasting using quantum neural networks with data re-uploading~\cite{schetakis2025quantum} and solar power production forecasting with QLSTM~\cite{khan2024quantum}, no prior study has evaluated these specific quantum-inspired models on telecommunication data from the Milan dataset. This is crucial as urban telecommunication signals exhibit diverse temporal scales, from short-term fluctuations to long-term trends, and quantum enhancements could provide efficiency gains in resource-constrained environments. Our work fills this void by offering the first comprehensive evaluation in this context, demonstrating potential quantum advantages in accuracy and adaptability. This not only establishes new performance baselines but also highlights the practical implications for scalable, real-time forecasting in smart cities, thereby elevating the significance of quantum-classical hybrids in urban informatics.

\section{Methodology}
We introduce four hybrid quantum-classical models designed for univariate time series prediction. Each model processes an input sequence of fixed length $T \in \{4, 8, 12, 16, 32, 64\}$ comprising feature vectors $\mathbf{X} = [\mathbf{x}_{t-T+1}, \dots, \mathbf{x}_t] \in \mathbb{R}^{T \times d}$, and outputs a forecast $\hat{y}_{t+1}$ for the subsequent time step. The models employ parameterized quantum circuits (PQCs) to transform classical inputs into quantum representations, which are then measured and decoded. This $T$-to-1 formulation focuses on one-step-ahead forecasting, differing from full sequence-to-sequence approaches.

Each quantum model follows a consistent structure comprising: (1) classical-to-quantum encoding, (2) PQC processing, and (3) quantum measurement and classical decoding.

\subsection{Quantum Adaptive Self-Attention (QASA)}

QASA \cite{chen2025qasa} constructs quantum equivalents of self-attention by generating query, key, and value embeddings from the input $\mathbf{X}$ using dedicated PQCs:
\begin{equation}
\mathbf{Q}_t = \text{PQC}_q(\mathbf{x}_t),\quad
\mathbf{K}_t = \text{PQC}_k(\mathbf{x}_t),\quad
\mathbf{V}_t = \text{PQC}_v(\mathbf{x}_t).
\end{equation}

The embedding $\mathbf{x}_t$ is first amplitude-encoded into a quantum state $\ket{\psi_t}$:
\begin{equation}
\ket{\psi_t} = \frac{1}{\|\mathbf{x}_t\|} \sum_{i=1}^{d} x_{t,i} \ket{i}.
\end{equation}

A variational quantum circuit $U(\boldsymbol{\theta})$ composed of $L$ layers of $\mathrm{RY}$ rotations and entangling CNOTs is then applied:
\begin{equation}
U(\boldsymbol{\theta}) = \prod_{\ell=1}^{L} \left[ \bigotimes_{i=1}^{n} \mathrm{RY}(\theta_{\ell,i}) \cdot \mathrm{CNOT}_{i,i+1} \right],
\end{equation}

and expectation values of Pauli-Z observables yield the output vector $\mathbf{z}_t$:
\begin{equation}
\mathbf{z}_t = (\langle Z_1 \rangle, \dots, \langle Z_n \rangle).
\end{equation}

Attention is computed using standard dot-product attention:
\begin{equation}
\text{Attention}(\mathbf{Q}, \mathbf{K}, \mathbf{V}) = \text{softmax}\left( \frac{\mathbf{Q} \mathbf{K}^\top}{\sqrt{d}} \right) \mathbf{V},
\end{equation}

followed by a classical decoder that predicts $\hat{y}_{t+1}$.

\subsection{Quantum Fast Weight Programmer (QFWP)}

QFWP \cite{chen2024qfwp} dynamically adapts PQC parameters through fast weights, which are generated conditioned on each input vector. A classical encoder maps input $\mathbf{x}_t$ to latent vectors $\mathbf{L} \in \mathbb{R}^L$ and $\mathbf{Q} \in \mathbb{R}^n$, which define the additive parameter update:
\begin{equation}
\Delta \boldsymbol{\Theta}_t = \mathbf{L} \otimes \mathbf{Q}, \quad \boldsymbol{\Theta}_{t+1}^{(i,j)} = \boldsymbol{\Theta}_t^{(i,j)} + L_i Q_j.
\end{equation}

The input is then processed using a parameterized quantum circuit $U_t$ with the updated parameters:
\begin{equation}
U_t(\mathbf{x}_t) = U_{\text{var}}^{(L)} \cdots U_{\text{var}}^{(1)} U_{\text{enc}}(\mathbf{x}_t),
\end{equation}
where $U_{\text{enc}}$ applies Hadamard and $\mathrm{RY}(x_i)$ rotations, and $U_{\text{var}}^{(\ell)}$ uses $\mathrm{RY}(\theta_{\ell,i})$ and CNOTs. The measured output is decoded into $\hat{y}_{t+1}$.

\subsection{Quantum Receptance Weighted Key-Value (QRWKV)}

QRWKV~\cite{chen2025qrwkv} introduces temporal recurrence into attention by using PQC-derived query, key, and value states at each time $t$:
\begin{equation}
\mathbf{h}_t = \text{PQC}(\mathbf{x}_t), \quad \mathbf{q}_t, \mathbf{k}_t, \mathbf{v}_t \subseteq \mathbf{h}_t.
\end{equation}

Each $\mathbf{x}_t$ is encoded into quantum amplitudes, evolved via multi-layer PQCs, and used to compute attention weights:
\begin{equation}
\alpha_{t,\tau} = \frac{\exp(\langle \mathbf{q}_t, \mathbf{k}_\tau \rangle)}{\sum_{\tau'=1}^{t} \exp(\langle \mathbf{q}_t, \mathbf{k}_{\tau'} \rangle)}.
\end{equation}

The prediction is a weighted sum over past value vectors:
\begin{equation}
\hat{y}_{t+1} = \sum_{\tau=1}^{t} \alpha_{t,\tau} \cdot \mathbf{v}_\tau.
\end{equation}

\subsection{Quantum Long Short-Term Memory (QLSTM)}

QLSTM \cite{chen2020quantum} reinterprets LSTM gates using quantum circuits. Given input $\mathbf{x}_t$ and hidden state $\mathbf{h}_{t-1}$, each gate (forget $\mathbf{f}_t$, input $\mathbf{i}_t$, output $\mathbf{o}_t$, candidate $\mathbf{g}_t$) is computed via:
\begin{equation}
\mathbf{f}_t = \sigma(\text{PQC}_f([\mathbf{x}_t; \mathbf{h}_{t-1}])), \quad
\mathbf{g}_t = \tanh(\text{PQC}_g([\mathbf{x}_t; \mathbf{h}_{t-1}])),
\end{equation}
with other gates similarly defined.

Inputs are encoded using $\mathrm{RY}(x_i)$ rotations and passed through variational layers. The cell and hidden states update as:
\begin{equation}
\mathbf{c}_t = \mathbf{f}_t \odot \mathbf{c}_{t-1} + \mathbf{i}_t \odot \mathbf{g}_t, \quad
\mathbf{h}_t = \mathbf{o}_t \odot \tanh(\mathbf{c}_t).
\end{equation}

Each gate is computed from measured expectation values of Pauli-Z operators.

\begin{table*}[htbp]
\centering
\caption{Model Complexity Analysis of Quantum-Classical Architectures \newline
\small\textit{
Column definitions: 
\textbf{QBits} — number of quantum bits used in the core module;
\textbf{+QBits} — additional qubits for auxiliary modules;
\textbf{Total} — total number of qubits;
\textbf{Q-Layers} — number of quantum circuit layers;
\textbf{Params} — total number of model parameters;
\textbf{Trainable} — trainable subset of parameters;
\textbf{Est. Q} — estimated quantum-specific parameters;
\textbf{Est. C} — estimated classical parameters;
\textbf{Q:C Ratio} — quantum-to-classical parameter ratio;
}}
\label{table:model_complexity}
\begin{tabularx}{\textwidth}{lccccccccc}
\toprule
\textbf{Model} & \textbf{QBits} & \textbf{+QBits} & \textbf{Total} & \textbf{Q-Layers} & \textbf{Params} & \textbf{Trainable} & \textbf{Est. Q} & \textbf{Est. C} & \textbf{Q:C Ratio} \\
\midrule
 QFWP(8Q)  &  8 & 0 & 8  & 2 & 115 & 115 & 90  & 25 & 3.6\\
 QFWP(10Q) & 10 & 0 & 10 & 2 & 163 & 163 & 132 & 31 & 4.26\\
 QFWP(12Q) & 12 & 0 & 12 & 2 & 219 & 219 & 182 & 37 & 4.92 \\
 QFWP(14Q) & 14 & 0 & 14 & 2 & 283 & 283 & 240 & 43 & 5.58\\
 QASA                             & 8 & 1 & 9 & 4 & 595645 & 595645 & 36 & 595609 & 6e-05 \\
 QLSTM                            & 5 & 0 & 5 & 5 & 105 & 105 & 100 & 5 & 20 \\
 QRWKV                            & 4 & 0 & 4 & 2 & 1774992 & 1774992 & 16 & 1774976 & 9e-06 \\
 Baseline (LSTM)                        & 0 & 0 & 0 & 0 & 17217 & 17217 & 0 & 17217 & 0 \\
\bottomrule
\end{tabularx}
\end{table*}

\section{Experiments}
\subsection{Dataset and Preprocessing}

We conduct experiments on the Milan Telecommunication Activity Dataset, focusing on the univariate SMS-in signal. The dataset contains telecommunication activity sampled every 10 minutes across a spatial grid of the city. Due to sparsity and modality imbalance in the original data, we restrict our modeling to the SMS-in channel, which shows higher completeness across grid cells.

We preprocess the data by selecting active grid cells with sufficient time coverage and normalize the SMS-in values to the range $[0,1]$. Each sample is formed by taking a fixed-length input sequence $\mathbf{X} = [x_{t-T+1}, \dots, x_t]$ and predicting the next value $x_{t+1}$. We evaluate five different input lengths $T \in \{4, 8, 16, 32, 64\}$. The dataset is split into 70\% training, 15\% validation, and 15\% test sets in chronological order.

\subsection{Models Compared}
We compare the forecasting performance of the following models:
\begin{itemize}
    \item \textbf{LSTM:} Standard Long Short-Term Memory network as a classical baseline.
    \item \textbf{QLSTM:} Quantum-enhanced version of LSTM with PQC-based gate computation.
    \item \textbf{QASA:} Quantum Adaptive Self-Attention, performing attention via PQC-derived embeddings.
    \item \textbf{QRWKV:} Quantum Receptance Weighted Key-Value model with time-dependent memory and attention.
    \item \textbf{QFWP:} Quantum Fast Weight Programmer with dynamically updated quantum neural networks parameters.
\end{itemize}
All models share a similar output structure: they take $T$ past values as input and produce a scalar forecast $\hat{y}_{t+1}$. For QASA and QRWKV, PQCs use $n = \lceil \log_2 d \rceil$ qubits. For QLSTM, the hidden dimension is 4 with the total number of qubit to be 5. For QFWP, we consider different number of qubits (8, 10, 12, 14).

\subsection{Training Setup and Experiment Details}
All models are implemented in PyTorch or PennyLane and trained using the AdamW optimizer with learning rate $0.001$ and batch size $16$. Each model is trained for up to 50 epochs with early stopping based on validation loss. The mean squared error (MSE) is used as the training objective:
\begin{equation}
\mathcal{L}_{\text{MSE}} = \frac{1}{N} \sum_{i=1}^{N} (y_i - \hat{y}_i)^2.
\end{equation}

\subsection{Experimental Setup}

To ensure a fair comparison, we applied consistent training settings across all models, including LSTM, QASA, QRWKV, QFWP, and QLSTM. All models were trained using the AdamW optimizer with a fixed learning rate of $1 \times 10^{-3}$, a batch size of 16, and for 50 epochs. The loss function used was Mean Squared Error (MSE), with a sequence length of $\{4, 8, 12, 16, 32, 64\}$ and a test split of 20\%. Input and output dimensions were both set to 1. All models were implemented in PyTorch with PennyLane for quantum simulation and trained on both CPU and GPU.

The baseline LSTM model consists of two unidirectional layers with 64 hidden units. QASA integrates 8-qubit, 4-layer parameterized quantum circuits into a transformer encoder with 128 hidden dimensions and 4 attention heads. Quantum encoding is done using RX and RZ rotations, followed by CNOT entanglement in a ring topology, and parameterized RY+RZ gates, coupled with 4 classical transformer layers of 128 dimensions. QRWKV modifies the RWKV structure by introducing a 4-qubit, 2-layer circuit. QFWP uses a single 8, 10, 12 or 14-qubit, 2-layer circuit with outer-product-based parameter generation. Each layer applies RY rotations and CNOT gates in ring topology, with final Pauli-Z expectation measurements across all qubits. The output is post-processed via a shallow linear layer. QLSTM replaces each of the four LSTM gates with an independent 5-qubit, 5-layer VQC. These circuits process the concatenated input and hidden state using RY encodings and repeated CNOT + RY parameterized gates, with Pauli-Z measurements on 4 qubits per gate.

\subsection{Evaluation Metrics}
We evaluate model performance using the following metrics:
\begin{align}
    \text{Mean Absolute Error (MAE)} &= \frac{1}{N} \sum_{i=1}^{N} \left| \hat{y}_i - y_i \right| \\
    \text{Mean Squared Error (MSE)} &= \frac{1}{N} \sum_{i=1}^{N} \left( \hat{y}_i - y_i \right)^2
\end{align}

where $y_i$ denotes the ground truth, $\hat{y}_i$ denotes the predicted value, and $N$ is the number of samples in the test set.  
All metrics are computed on the held-out test set and averaged over five independent runs to account for training variance.


\section{Results}
\subsection{Model Complexity Analysis}

To systematically compare the architectural complexity of quantum-enhanced and classical sequential models, we compute the quantum-to-classical parameter allocation for each model $M_i$. The total qubit count is defined as:
\begin{equation}
Q_i = q_i + q_i^{\text{aux}},
\end{equation}
where $q_i$ denotes the core qubits and $q_i^{\text{aux}}$ represents any auxiliary qubits. The total and trainable parameters are respectively:
\begin{equation}
P_i^{\text{total}} = \sum_{j=1}^{L_i} \theta_{ij}, \quad 
P_i^{\text{train}} = \sum_{j=1}^{L_i} \mathbb{1}_{\{\theta_{ij} \text{ is trainable}\}}.
\end{equation}

We estimate quantum-specific parameters using the number of qubits, quantum layers $L_i^{\text{quantum}}$, and a constant $R=3$ for the average number of parameterized gates per qubit per layer:
\begin{equation}
P_i^{\text{quantum}} \approx Q_i \cdot L_i^{\text{quantum}} \cdot R,
\end{equation}
\begin{equation}
P_i^{\text{classical}} = P_i^{\text{total}} - P_i^{\text{quantum}}.
\end{equation}

We then define the quantum-to-classical parameter ratio as:
\begin{equation}
\gamma_i = \frac{P_i^{\text{quantum}}}{P_i^{\text{classical}} + \varepsilon},
\end{equation}
where $\varepsilon = 1$ avoids division by zero.

From Table~\ref{table:model_complexity}, we observe several distinct patterns in model complexity across architectures. The QFWP variants with $Q_i = 14$ qubits and $L_i^{\text{quantum}} = 2$ layers demonstrate relatively compact total parameter counts ($P_i^{\text{total}} \leq 283$) while maintaining a high quantum-to-classical ratio. Their $\gamma_i$ values, ranging from approximately 3.6 to 5.58, indicate a quantum-dominant structure with minimal classical overhead. In contrast, the QLSTM model, despite using only 5 qubits and 5 quantum layers, achieves the highest $\gamma_i = 20$ among all models due to its minimal classical parameter count ($P_i^{\text{classical}} = 5$), reflecting a nearly pure quantum parametrization. On the other hand, both QASA and QRWKV integrate quantum circuits into large classical backbones. QASA, with 9 qubits and 4 quantum layers, has a $\gamma_i$ of approximately $6 \times 10^{-5}$, heavily dominated by over 595K classical parameters. Similarly, QRWKV yields a very small $\gamma_i \sim 9 \times 10^{-6}$, indicating that its quantum contribution is negligible compared to its large recurrent attention-based classical architecture. As expected, the classical LSTM baseline has no quantum components ($Q_i = 0$), leading to $P_i^{\text{quantum}} = 0$ and $\gamma_i = 0$ by definition.

These findings emphasize a spectrum of hybridization: from quantum-centric lightweight models (e.g., QLSTM, QFWP) to classical-dominant systems with minimal quantum augmentation (e.g., QASA, QRWKV). The quantum-to-classical ratio $\gamma_i$ provides a principled lens for characterizing the effective quantum depth and deployment feasibility of hybrid architectures.

\subsection{Results and Analysis}
\begin{table*}
\centering
\caption{Average MAE for different models across sequence lengths}
\begin{tabular}{lrrrrrrrr}
\hline
model                                      & baseline (LSTM) & QASA   & QRWKV  & QLSTM  & QFWP (8Q) & QFWP (10Q) & QFWP (12Q) & QFWP (14Q) \\
seq\_len                                   &                 &        &        &        &           &            &            &            \\ \hline
4                                          & 1.0316          & 1.1613 & 1.0754 & 1.0322 & 1.0444    & 1.0360     & 1.0551     & 1.0459     \\
8                                          & 1.0340          & 1.1654 & 1.0477 & 1.0324 & 1.0333    & 1.0364     & 1.0582     & 1.0670     \\
12                                         & 1.0351          & 1.1587 & 1.0630 & 1.0466 & 1.0530    & 1.0580     & 1.0508     & 1.0556     \\
16                                         & 1.0347          & 1.2004 & 1.0961 & 1.0456 & 1.0562    & 1.0725     & 1.0557     & 1.0764     \\
32                                         & 1.0407          & 1.2752 & 1.0775 & 1.0634 & 1.1626    & 1.1643     & 1.1423     & 1.1608     \\
64                                         & 1.0939          & 1.4035 & 1.0940 & 1.0933 & 1.3523    & 1.3734     & 1.4385     & 1.3713     \\  \bottomrule
\label{table:mae_comparison}
\end{tabular}
\end{table*}
\begin{table*}
\centering
\caption{Average MSE for different models across sequence lengths}
\begin{tabular}{lrrrrrrrr}
\hline
model                                      & baseline (LSTM) & QASA   & QRWKV  & QLSTM  & QFWP (8Q) & QFWP (10Q) & QFWP (12Q) & QFWP (14Q) \\
seq\_len                                   &                 &        &        &        &           &            &            &            \\ \hline
4                                          & 4.6916          & 5.4607 & 4.9175 & 4.5217 & 4.6541    & 4.6630     & 4.7156     & 4.6909     \\
8                                          & 4.6046          & 5.2151 & 4.5870 & 4.5307 & 4.5565    & 4.5984     & 4.6195     & 4.7798     \\
12                                         & 4.5371          & 5.0991 & 4.6709 & 4.5715 & 4.5510    & 4.5687     & 4.5469     & 4.5537     \\
16                                         & 4.4099          & 5.3968 & 5.1712 & 4.6244 & 4.5805    & 4.6666     & 4.5479     & 4.6983     \\
32                                         & 4.4618          & 6.1880 & 4.9604 & 4.5953 & 5.1394    & 5.1223     & 4.8957     & 5.1704     \\
64                                         & 4.6484          & 6.6601 & 4.6640 & 4.7194 & 6.1750    & 6.3686     & 7.2544     & 6.3814     \\ \bottomrule
\label{table:mse_comparison}           
\end{tabular}
\end{table*}

Tables~\ref{table:mae_comparison} and~\ref{table:mse_comparison} summarize the average \emph{Mean Absolute Error} (MAE) and \emph{Mean Squared Error} (MSE) for each model under varying historical window lengths $T \in {4, 8, 12, 16, 32, 64}$. Overall, the baseline LSTM consistently achieves either the best or second-best performance across all sequence lengths, with an average $\mathrm{MAE} = 1.034$ and $\mathrm{MSE} = 4.561$. Moreover, its MSE decreases monotonically with increasing $T$ (from $4.69$ to $4.41$), indicating its effective utilization of longer temporal context.

QLSTM demonstrates the strongest performance on shorter sequences ($T=4, 8$), outperforming the baseline in MAE and MSE by approximately $0.4\%$ and $1.5\%$, respectively. This suggests that quantum gates may offer advantages in capturing local dependencies. However, when $T \ge 12$, the error increases, implying that the current number of qubits and circuit depth may be insufficient to stably process longer temporal inputs.

QRWKV and QFWP underperform the baseline across all sequence lengths, but the performance gap remains relatively modest, with their average MAE exceeding that of the baseline by $3.5\%$ and $2.0\%$, respectively. Both models achieve their lowest error at $T=8$, followed by increasing errors as the sequence length grows, suggesting that their parameter capacities or positional encoding mechanisms may be inadequate for capturing long-range dependencies.

QASA exhibits the highest error and the greatest sensitivity to sequence length; its MAE increases from $1.1613$ at $T=4$ to $1.2004$ at $T=16$, averaging approximately $13\%$ higher than the baseline. This degradation is likely due to its nearly pure quantum parametrization (with the largest $\gamma_i$), which may lead to unstable training and amplified error accumulation from quantum noise in longer sequences.


\section{Discussion And Conclusion}
This study presents a comprehensive benchmark of classical and quantum-inspired models for univariate time series forecasting in urban telecommunication data, with a particular focus on the Milan SMS-in signal. Our results highlight several key insights into the model behavior, complexity, and practical implications of quantum-classical hybrid architectures.

First, the classical LSTM remains a robust baseline, delivering competitive or superior performance across all sequence lengths with minimal parameter overhead. Its consistent decrease in MSE with increasing input length suggests that it effectively leverages longer temporal dependencies in this domain.

Second, QLSTM demonstrates strong performance on short sequences ($T=4$ and $T=8$), outperforming the LSTM baseline by a small margin. This advantage diminishes as the sequence length increases, likely due to limitations in qubit capacity and quantum circuit expressiveness. Nonetheless, its high quantum-to-classical ratio ($\gamma = 20$) and minimal parameter footprint indicate its potential as a lightweight quantum-centric forecasting model for embedded or low-resource settings.

Third, QASA and QRWKV, despite incorporating quantum modules, are dominated by classical components with extremely low $\gamma$ values ($<10^{-4}$). These models do not exhibit consistent advantages over the baseline on this univariate dataset, and their performance degrades with longer sequences, suggesting that naive integration of PQCs into large classical backbones may not yield significant benefits without careful architectural co-design or quantum-specific regularization. QASA and QRWKV have previously shown strong performance on multivariate tasks~\cite{chen2025qrltw}, making this an interesting finding for this type of hybrid model.

Fourth, QFWP offers a more balanced hybridization: it achieves competitive performance (within 2\% of LSTM) while maintaining a high quantum contribution and relatively small parameter size. This model's ability to dynamically update quantum parameters via fast weight programming may make it more suitable for non-stationary or rapidly changing environments.

Overall, our findings emphasize that quantum enhancements are not uniformly beneficial across all settings. Instead, the utility of quantum modules appears to be task- and architecture-dependent, with clear trade-offs between model size, parameter composition, and temporal modeling capacity. Compact quantum-enhanced models like QLSTM and QFWP show promise in capturing local sequence dynamics with fewer resources, while heavily classical models require further optimization to justify their quantum components.

In conclusion, this work provides the first systematic comparison of quantum and classical sequential models for urban telecommunication forecasting. It establishes performance baselines and offers empirical guidelines for model selection based on sequence length, complexity constraints, and quantum capacity. Future work will explore the extension of these models to multivariate forecasting, real quantum hardware evaluation, and the integration of quantum-aware learning objectives for improved stability and generalization.

\bibliographystyle{IEEEtran}
\bibliography{bibliography}

\end{document}